\documentclass[a4paper,12pt]{article}

\usepackage{epsfig}
\usepackage{amstex}
\usepackage{amssymb}

\def\as{\alpha_s}
\def\ms{m_{\tilde q}}
\def\mg{m_{\tilde g}}

\def\gl{\tilde{g}}
\def\sq{\tilde{q}}
\def\sqb{\bar{\tilde{q}}}
\def\DR{$\overline{DR}$}
\def\MS{$\overline{MS}$}
\def\ghat{\hat{g}_s}
\def\Li{\text{Li}_2}

\begin{document}
\thispagestyle{empty}

\hfill\vbox{\hbox{\bf DESY 96-022}
            \hbox{February 1996}
                                }
\vspace{1.in}
\begin{center}
\renewcommand{\thefootnote}{\fnsymbol{footnote}}
{\large\bf SUSY-QCD Decays of Squarks and Gluinos}\\
\vspace{0.5in}
W.~Beenakker$^{1}$\footnote{Research supported by a fellowship of the 
Royal Dutch Academy of Arts and Sciences}, R.~H\"opker$^2$, 
and P.~M.~Zerwas$^2$ \\ 
\vspace{0.5in}
$^1$ Instituut-Lorentz, University of Leiden, The Netherlands\\
$^2$ Deutsches Elektronen-Synchrotron DESY, D-22603 Hamburg, Germany \\
\end{center}
\vspace{5cm}

\begin{center}
ABSTRACT \\
\end{center}

The partial widths are determined for squark decays to gluinos and
quarks, and gluino decays to squarks and quarks, respectively.  The
widths are calculated including one-loop SUSY-QCD corrections. The
corrections amount to $+$30\% to $+$50\% for squark decays and
$-$10\% to $+$10\% for gluino decays. We have derived the
results in the \DR ~and \MS ~renormalization schemes, and we have
demonstrated explicitly that the one-loop effective $qqg$ and
$q\sq\gl$ couplings are equal in the limit of exact supersymmetry.

\pagebreak
\setcounter{footnote}{0}

\section{Introduction}

The theoretical predictions for the decay properties of squarks and
gluinos in supersymmetric theories are of interest for several
reasons. The decay properties determine the experimental search
techniques, and the measurement of decay branching ratios will allow
us to study the couplings between the standard and the novel
particles, interrelated by supersymmetry. A thorough analysis of the
partial widths including higher-order corrections is required for this
purpose at the theoretical level. SUSY-QCD corrections, which take
into account the spectrum of the standard quarks/gluons and their
supersymmetric partners squarks/gluinos, have so far been performed
only for a few examples: The production of squark and gluino pairs at
hadron colliders \cite{BHSZ} and of squark pairs in $e^+e^-$
annihilation \cite{abdel}, the decay of squarks to quarks and photinos
\cite{hikasa}, and the decay of charged Higgs bosons to stop/sbottom
\cite{bartl1} and top/bottom particles \cite{bartl2}. In this note we
present the theoretical predictions for the decay channels
\begin{eqnarray}
  \sq &\to& \gl + q \hspace{2.2cm}\text{for}\quad \ms > \mg
  \label{process1}\\
  \gl &\to& \sq + \bar{q} ~/~ \sqb+q \qquad\text{for}\quad \mg > \ms
  \label{process2}
\end{eqnarray}
including SUSY-QCD corrections. The analysis is performed for light
quarks, excluding the case of top (s)quarks. This calculation is more
demanding than the calculation for the decay processes discussed
earlier in the literature due to the three colored particles involved
in the initial and final states of (\ref{process1}) and
(\ref{process2}).  At the technical level, we will follow the path
traced out for the production of squarks and gluinos in hadron
collisions \cite{BHSZ}.

\section{Theoretical Set-up}
\label{set-up}

In analyzing the SUSY-QCD corrections, the supersymmetric squarks are
assumed to be degenerate in mass. The formulae for squark decays apply
to both chiral states, $\sq_L$ and $\sq_R$. For gluino decays they
include the sum over all flavor and chiral components and the two
possible final states in Eq.~(\ref{process2}) related by ${\cal C}$
conjugation.  The final states involving top quarks are not taken into
account since we have treated the quarks in the final states as
massless particles; the case of (s)top particles will be deferred to a
subsequent analysis. In loops we have used a top-quark mass of
$m_t=180$~GeV and we have taken the top-squarks to be mass-degenerate
with the other squarks. The masses of the squarks and gluinos are
assumed to be above 100~GeV. (For experimental bounds see
Ref.~\cite{susyexp}). The squark and gluino masses are defined as pole
masses.  \medskip

\begin{figure}[t]
  \begin{center}
    \vspace*{-2.5cm}
    \epsfig{file=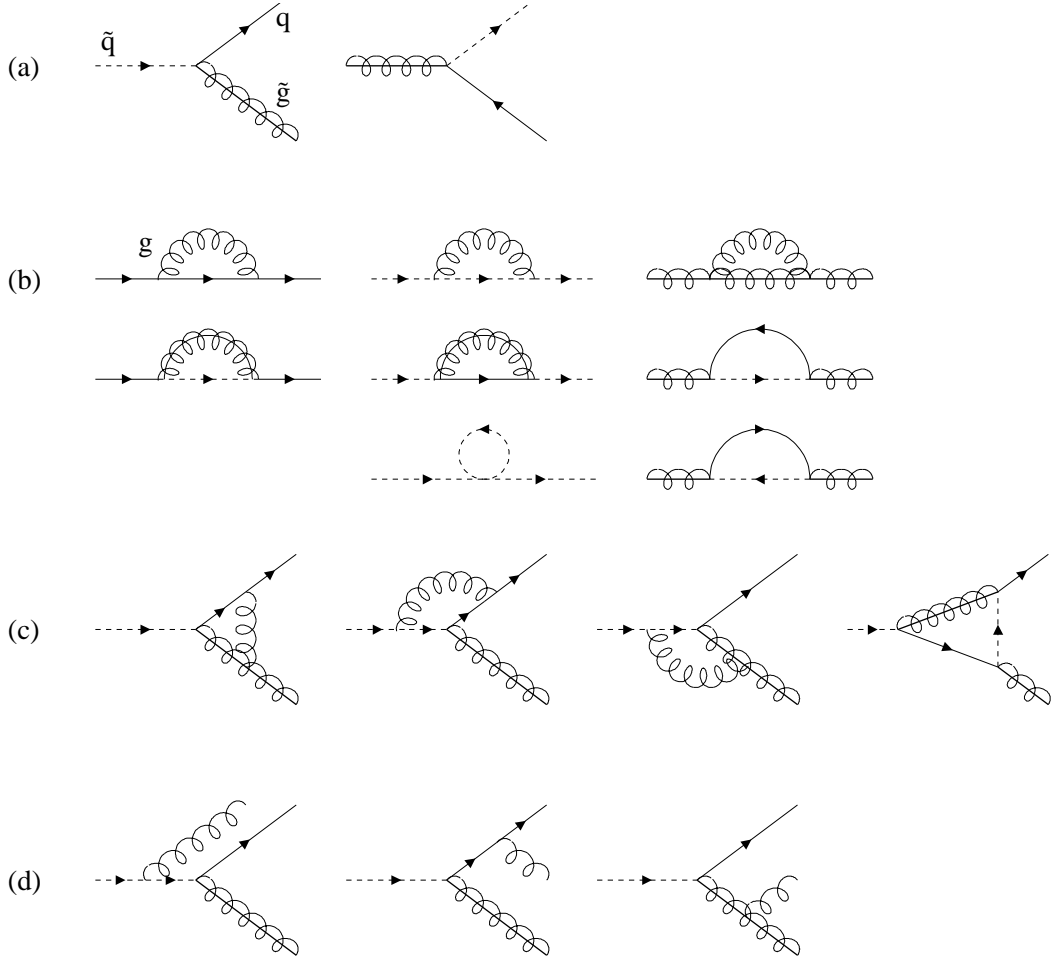,width=16cm}
  \end{center}
  \caption{The Feynman diagrams for squark and gluino decays. 
    (a) Born diagram for both decay modes;
    (b) quark, squark, and gluino self-energies;
    (c) vertex corrections for squark decays; 
    (d) gluon radiation for squark decays. 
    The diagrams of type (c) and (d) for gluino decays can be
    generated by rotating the diagrams for squark decays.}
  \label{feyn}
\end{figure}

The diagrams for the squark and gluino decays are displayed in
Fig.~\ref{feyn}. While (a) represents the Born diagrams for the two
decay modes, the self-energy diagrams are shown in (b), vertex
corrections in (c), and hard-gluon radiation in (d). In (c) and (d) we
only give the squark-decay diagrams, as the gluino and squark decay
processes are related to each other by crossing.

\bigskip
The ultra-violet, infrared, and collinear divergences have been regularized 
by switching to $n=4+\varepsilon$ dimensions. The actual calculation has been
carried out in two schemes:

(i) It is well-known that dimensional reduction \DR ~\cite{siegel}
preserves supersymmetry at least to one-loop. This scheme is hence an
obvious candidate for performing the renormalization; some technical
difficulties in handling the factorization of collinear singularities,
however, still demand a proper understanding \cite{neerven}.

(ii) From a technical point of view, the \MS ~renormalization
scheme \cite{buras} is easier to handle in general. However, by
introducing a mismatch between the number of gauge-boson and gaugino
degrees of freedom in $n\neq 4$ dimensions, the scheme violates
supersymmetry explicitly. In particular, the $q\sq\gl$ Yukawa coupling
$\ghat$, which by supersymmetry should coincide with the $qqg$ gauge
coupling $g_s$, deviates from $g_s$ by a finite amount at the one-loop
order. Requiring the physical amplitudes to be independent of
the renormalization scheme, a shift between the bare Yukawa and gauge
couplings must be introduced in the \MS ~scheme \cite{martin},
\begin{equation}
  \hat{g}_s = g_s \left[ 1 + \frac{\as}{4\pi}\left(\frac{2}{3}C_A
  -\frac{1}{2}C_F\right)\right]
  \label{finshift}
\end{equation}
which effectively subtracts the contributions of the false,
non-supersymmetric degrees of freedom (also called $\varepsilon$
scalars); $C_A=3$ and $C_F=4/3$ are the Casimir invariants of the
$SU(3)$ gauge group.

\bigskip Taking into account this shift in the \MS ~scheme, we find
that all results are identical with the results obtained within the
\DR ~scheme bearing in mind the relation between the gauge couplings
in the two schemes.  This comparison provides non-trivial consistency
checks of the one-loop SUSY-QCD corrections.

\medskip
The need for introducing a finite shift is best demonstrated for
the effective $q\sq\gl$ Yukawa coupling, which must be equal to the
$qqg$ gauge coupling in a supersymmetric world with massless
gauginos/gluons and equal-mass squarks/quarks. For the sake of
simplicity we define the effective couplings $g_s^{eff}(Q^2)$ and
$\hat{g}_s^{eff}(Q^2)$ in the limit of on-shell squarks/quarks
and almost on-shell gluinos/gluons with virtuality $Q^2 \ll \ms^2 =m_q^2$. 
[In this limit the couplings do not involve gauge-dependent terms.]

\medskip
First we consider the corrections of the couplings
\begin{equation}
  g_s^{eff}(Q^2) = g_s (1 + \delta) \quad\text{and}\quad
  \ghat^{eff}(Q^2) = \ghat (1 +\hat{\delta})
\label{gseff}
\end{equation}
in the \MS ~scheme. By adding the contributions of the external
self-energies and the charge renormalization to the vertex corrections
[all given explicitly in the appendix], 
we find the following
results\footnote{The remaining $1/\varepsilon$ poles are associated
  solely with infrared and collinear singularities, while all ultra-violet
  singularities are cancelled.}:
\begin{eqnarray}
  \overline{MS}:\quad \hat{\delta}(q\sq\gl) & = &
  \frac{\as}{4\pi}\,\left\{ C_A\,\left[ \frac{2}{\varepsilon} -
  \frac{1}{2}\,\log\left(\frac{Q^2}{\ms^2}\right) + \frac{1}{2}
\right] + \frac{C_F}{2} \right\} \label{qsqgltot}\\[2mm] 
\delta(qqg) & = & \frac{\as}{4\pi}\,C_A\,\left[
\frac{2}{\varepsilon} -
\frac{1}{2}\,\log\left(\frac{Q^2}{\ms^2}\right) + \frac{7}{6} \right]
      \label{qqgtot}
\end{eqnarray}
The difference between $\delta$ and $\hat{\delta}$ is indeed
compensated by the finite shift introduced in Eq.~(\ref{finshift}). As
a result, the couplings $g_s^{eff}(Q^2)$ and $\hat{g}^{eff}_s(Q^2)$
agree with each other to one-loop, as required by supersymmetry. The
same observation can be made in the \DR ~scheme. In this scheme we
find
\begin{eqnarray}
  \overline{DR}:\quad \hat{\delta}(q\sq\gl) &= &
  \frac{\as}{4\pi}\,C_A\,\left[ \frac{2}{\varepsilon} -
  \frac{1}{2}\,\log\left(\frac{Q^2}{\ms^2}\right) + 1
\right] \\[2mm] \delta(qqg) &= &
\frac{\as}{4\pi}\,C_A\,\left[ \frac{2}{\varepsilon} -
\frac{1}{2}\,\log\left(\frac{Q^2}{\ms^2}\right) + 1 \right]
\end{eqnarray}
As anticipated, the \DR ~renormalization scheme preserves supersymmetry and
guarantees that the gauge and Yukawa couplings also coincide to one-loop order
in the supersymmetric limit. (Note the
difference between the effective gauge couplings in the \MS ~and \DR
~scheme, $(\as/4\pi)\times C_A/6$; see also Ref.~\cite{martin,schuler}).

\section{Results}
At the Born level, the partial widths of the decay processes (\ref{process1}) 
and (\ref{process2}) are given by
\begin{eqnarray}
  &&\Gamma_0(\sq\to\gl +q) = \frac{\as C_F}{2}\, 
    \frac{(\ms^2 -\mg^2)^2}{\ms^3}\\
  &&\Gamma_0(\gl\to\sqb +q,\sq+\bar{q}) = 
  4n_f\,\frac{\as}{8}\,\frac{(\mg^2 -\ms^2)^2}{\mg^3}
\end{eqnarray}
It is implicitly understood in the subsequent formulae, that all
flavors, helicities and ${\cal C}$ related modes for the gluino decays
are summed up. Adding up virtual corrections and gluon bremsstrahlung,
the final result including one-loop SUSY-QCD corrections may be
written in the form:
\begin{eqnarray}
  \!\!\Gamma(\sq) & = & \Gamma_0(\sq)\left\{ 1 +\frac{\as}{\pi}\,
  \big[ C_A F_A + C_F F_F + n_f F_f + F_t + F_{ren}
    +F_{dec}\big]\right\} \label{gnlosq}\\[2mm] 
  \!\!\Gamma(\gl) & = & \Gamma_0(\gl) \left\{ 1 
  +\frac{\as}{\pi}\,\big[ C_A (F_A-\pi^2)
  + C_F (F_F+\pi^2) + n_f F_f +F_t + F_{ren}
  + F_{dec}\big]\right\} \label{gnlogl}
\end{eqnarray}
The coefficients $F_A$ etc. are universal functions for both
processes, depending only on the respective mass ratios
\begin{displaymath}
  r = \frac{\mg^2}{\ms^2} \qquad\text{and}\qquad t = \frac{\mg^2}{m_t^2}
\end{displaymath}
The difference between the formulae (\ref{gnlosq}) and (\ref{gnlogl})
for the two processes is the 
result of the analytic continuation of the universal virtual
corrections from the region $\ms > \mg$ to the region $\mg > \ms$, which gives
rise to the $\pi^2$ terms. 
The coefficients read explicitly:                                
\begin{eqnarray*}
  F_A & = & \frac{3}{r-1}\,\Li(1-r) -\frac{r}{r-1}\,\Li(-r)
  +\frac{5r-6}{12(r-1)}\,\pi^2 +\frac{59}{24} +\frac{r}{4(r-1)}\\[1mm]
  & & {}+\left[\frac{3+r}{2(r-1)}\,\log(r) -2\right]\log|1-r| +
  \left[\frac{r(5r-6)}{4(r-1)^2}\, -\frac{r}{r-1}\, \log(1+r)\right]\log(r)
    \\[3mm]
  F_F & = & {}-\frac{2}{r-1}\,\Li(1-r) +\frac{2r}{r-1}\,\Li(-r)
  +\frac{4-3r}{6(r-1)}\,\pi^2 +\frac{5}{2} -\frac{r}{2} \\[1mm]
  & & {}+\left[r -\frac{r^2}{2}\,-\frac{r+1}{r-1}\,\log(r)\right]\log|1-r| 
  +\left[\frac{2r}{r-1}\,\log(1+r)-r+\frac{r^2}{2}\,\right]\log(r) \\[3mm]
  F_f &=& {}-\frac{3}{4r} +\frac{(r-1)(r+3)}{4r^2}\,\log|1-r|\\[3mm]
  F_t & =& \frac{1}{4r} -\frac{1}{4t}\left[ 1
  -\log\left(\frac{r}{t}\right) \right]   
  +\frac{1}{4}\left(\frac{1}{t}-\frac{1}{r} -1\right)B_0
  +\frac{1}{2}\left(\frac{1}{r}-\frac{1}{t}-1\right)B_0' 
\end{eqnarray*}
with
\begin{eqnarray*}
  B_0 &=& {\cal\text{Re}}\,\left[ \vphantom{\frac{1}{2}} 2 +x_1\log(1-1/x_1) 
          +x_2\log(1-1/x_2)\right]\\[1mm]
  B_0'&=& {\cal\text{Re}}\,\left[ -1 +\frac{
        x_1(1-x_1) \log(1-1/x_1) - x_2(1-x_2) \log(1-1/x_2) }{x_1
        -x_2} \right]\\[1mm]
  x_{1,2} & =& \frac{1}{2}\left[ 1 +\frac{1}{t} -\frac{1}{r} \pm
  \sqrt{(1+1/t -1/r)^2 -4/t} \,\right]
\end{eqnarray*}
The $F_F$ part represents the supersymmetric version of the abelian
vertex and self-energy contributions. The $F_f$ and $F_t$ contributions
are due to quark--squark self-energy 
corrections of the gluino; the gluino--gluon loop is incorporated in
$F_A$. In addition we have introduced the renormalization functions
\begin{eqnarray*}
  F_{ren} &=& \frac{3 C_A -n_f -1}{4}\,
  \log\left(\mu_R^2/\ms^2\right)\\[1mm]
  F_{dec} &=& \frac{n_f+1}{12}\,\log\left(\mu_R^2/\ms^2\right) 
  +\frac{1}{6}\,\log\left(\mu_R^2/m_t^2\right) 
  +\frac{C_A}{6}\,\log\left(\mu_R^2/\mg^2\right)
\end{eqnarray*}
where $\mu_R$ denotes the renormalization scale of the coupling $\as$.
The second part $F_{dec}$ is added to decouple the heavy particles (top
quarks, squarks, gluinos) from the running $\as$. [This term may be
omitted if the running coupling is adjusted properly].  

\begin{figure}[t]
  \begin{center} 
    \vspace*{-2.5cm}
    \epsfig{file=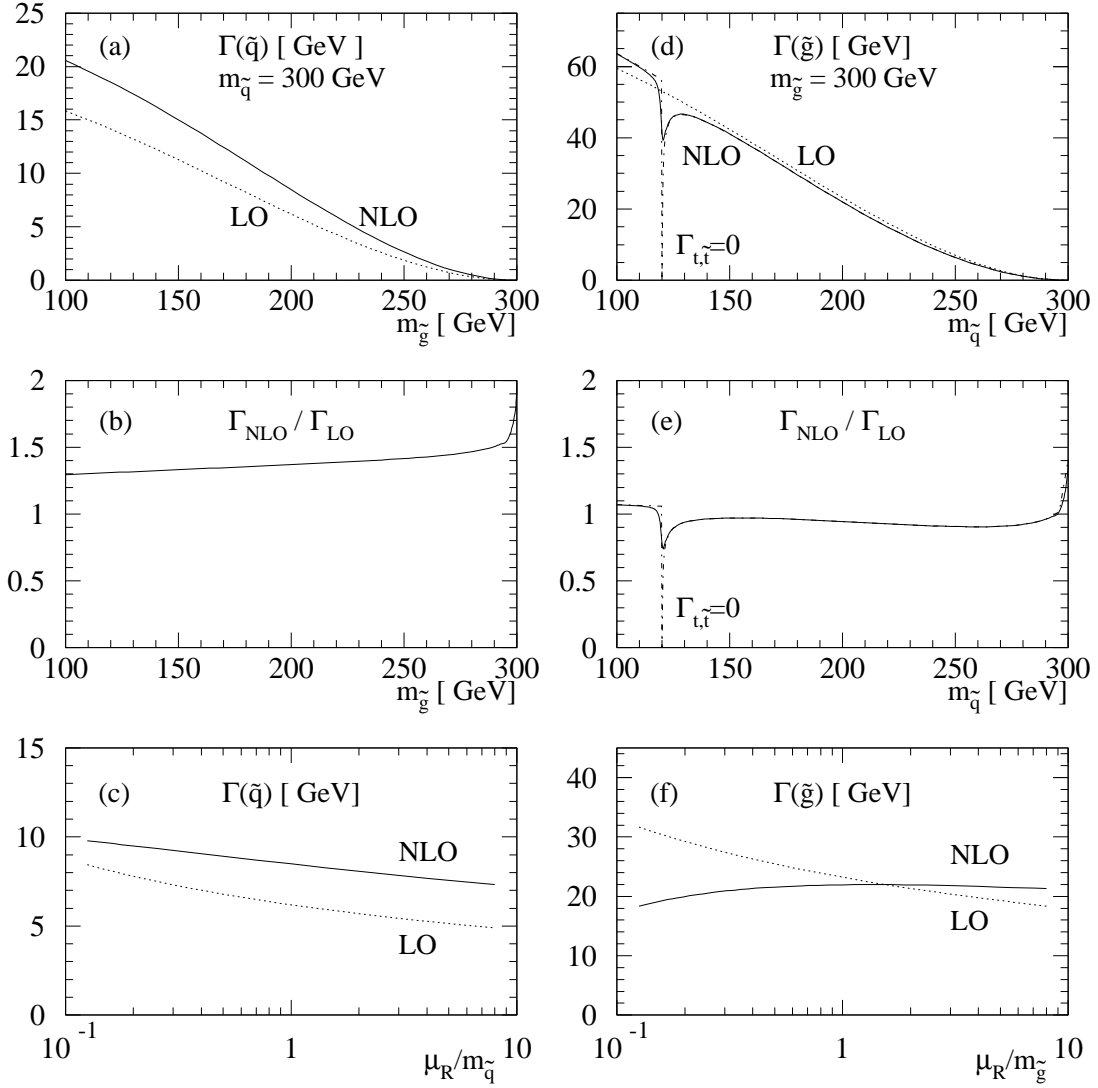,width=16cm}
  \end{center}
  \caption{The SUSY-QCD corrections to squark and gluino decays. 
    (a) The decay rate for squarks ($\ms=300$ GeV) in LO
        (dashed curve) and in NLO (solid curve) for $\mu_R=\ms$;
    (b) the ratio of the decay rate for squarks in NLO and in LO for 
        $\mu_R=\ms$;
    (c) the scale dependence of the decay rate for squarks in LO and in NLO 
        ($\mg=200$ GeV).
    (d-f): The same quantities for gluino decays with
    interchanged, but otherwise identical mass parameters. 
           The long-dashed curve in (d) and (e) represents NLO
           results without taking into account non-zero widths for stop
           and top.}
  \label{decayrates} 
\end{figure}

\bigskip Characteristic features of the corrections are exemplified in
Fig.~\ref{decayrates}. The three figures in the left column display
the width $\sq\to\gl q$, the figures in the right column the width
$\gl\to\sqb q,\sq \bar{q}$ for five light quark flavors.

\medskip The corrections for the squark decays are moderate. In
Fig.~\ref{decayrates}(a) the Born approximation (LO) and the
next-to-leading order (NLO) SUSY-QCD results are displayed for
$\mu_R=\ms$. The masses are chosen to be $\ms=300$~GeV, $\mg\ge
100$~GeV, and $m_t = 180$~GeV.  The running coupling $\as(\mu_R^2)$ is
normalized such that $\as(M_Z^2) = 0.118$. As outlined before, the
running of the coupling is defined in next-to-leading order by the
standard QCD spectrum with $n_f=5$ flavors, and the heavy particles
are decoupled. The size of the corrections\footnote{Since
  $\Lambda_{QCD}$ is not really well-defined in LO, we have identified
  $\as$ in LO with $\as$ in NLO at the mass scale chosen for
  comparison.}, measured by the ratio $\Gamma_{NLO}/\Gamma_{LO}$ in
Fig.~\ref{decayrates}(b), depends on the mass ratio $\mg/\ms$. With
corrections between +30\% and +50\%, they appear sizable but still
under control.

\medskip
The gluino decays are discussed, in a similar way, 
in Fig.~\ref{decayrates}(d) and (e). The corrections are considerably
smaller in this case.  The dip in the NLO curve is due to the
external gluino self-energy, related to the
$\gl\to\tilde{t}\bar{t}$ threshold in the stop--top loop.  The
singularities are artificial; they are smoothed out by the
strong stop--top interactions and the finite lifetimes of these
particles. Including the widths for $\tilde{t}$ and $\bar{t}$ in
$F_t$ removes part of the irregularity (see full curves). 
In view of the small
region in parameter space where this effect plays a role, we refrain
from a more detailed discussion.

\medskip
It is evident from the last figures in each column, Fig.~\ref{decayrates}(c) 
and (f), in
which the variation of the widths with the renormalization scale
$\mu_R$ is shown, that the NLO corrections improve the stability of
the theoretical predictions for the widths significantly. This is the case in
particular for the gluino decay width, where a wide and
shallow extremum is observed for $\mu_R/\mg$ near unity.

\bigskip To shed more light on these somewhat complicated results, it is
interesting to consider two limiting cases. We have considered the
massless limit for the particles in the final states, and the
threshold effects in the limit of nearly equal masses for squarks and
gluinos. For the sake of simplicity, we do not decouple the gluinos
and squarks from $\as$ and we take for illustration also the top mass equal
to zero. Setting the scale $\mu_R$ equal to the mass of the decaying particle,
we find:

\medskip
\noindent
$\underline{\text{\textit{massless gluinos/squarks in the final state:}}}$
\begin{eqnarray*}
  \Gamma(\sq ) &=& \Gamma_0(\sq)\left\{ 1+ \frac{\as}{\pi}\,
  \left[ \frac{59}{24}\,C_A +\left(\frac{5}{2}-\frac{\pi^2}{3}\right)
    C_F -\,\frac{n_f+1}{8} \right]\right\} \\[1mm] 
    \Gamma(\gl) &=& 
    \Gamma_0(\gl)\left\{ 1+ \frac{\as}{\pi}\,\left[
    \left(\frac{65}{24} - \frac{5 \pi^2}{12}\right)C_A + \left(\frac{7}{4}
    +\frac{\pi^2}{6}\right)C_F \right] \right\}
\end{eqnarray*}

\noindent
$\underline{\text{\textit{threshold limit $\ms \approx \mg$:}}}$
{\small
\begin{eqnarray*}
  \Gamma(\sq ) &=& \Gamma_0(\sq) \left\{ 1+\frac{\as}{\pi}\,
  \left[ \frac{3C_F}{2}\,\log\left(\frac{\ms^2}{\ms^2-\mg^2}\right)
    -\frac{3(n_f+1)}{4}+\frac{5 C_A}{6} +4 C_F +\frac{\pi^2 C_A}{2}
    -\frac{2 \pi^2 C_F}{3} \right] \right\} \\[1mm] 
    \Gamma(\gl) &=& \Gamma_0(\gl) 
    \left\{ 1+ \frac{\as}{\pi}\,\left[
    \frac{3C_F}{2}\,\log\left(\frac{\mg^2}{\mg^2-\ms^2}\right)
    -\frac{3(n_f+1)}{4}+\frac{5 C_A}{6} +4 C_F -\frac{\pi^2 C_A}{2}
    +\frac{\pi^2 C_F}{3} \right]\right\}
\end{eqnarray*}
} 
The different size of the corrections for squark and gluino decays
is nicely illustrated by the limiting case of massless particles in
the final state. The driving term for squark decays is the large $C_A$
contribution while other color structures play a minor role.  For
gluino decays, however, the different $\pi^2$ terms, induced by the
vertex correction for $\mg > \ms$, lead to a destructive interference
between the reduced (negative) $C_A$ contribution and the enhanced
(positive) contribution from the other color structures.  As a result
the SUSY-QCD corrections to gluino decays are significantly smaller
than the corrections for squark decays.

\bigskip In addition to the proof that the effective Yukawa and gauge
couplings are identical in the limit of exact supersymmetry, we have
performed several other consistency checks. We have shown, by using a
small gluon mass in the \DR ~scheme, that infrared and final-state
collinear divergences do not give rise to finite differences between
the \MS ~and the \DR ~schemes. The limiting case $\mg\to 0$ discussed
in the previous paragraph has been extracted from the general formulae
and it has been compared with an \emph{ab initio} rederivation of the
squark width (which can be calculated much more easily in this limit).
Finally, by limiting the calculation to the subset of abelian diagrams
(\emph{i.e.}~the $C_F$ part), we have reproduced the results of
Ref.~\cite{hikasa} for SUSY-QCD corrected squark decays to quarks and
photinos after having corrected the sign of the vertex correction Eq.~(19)
in Ref.~\cite{hikasa}.

\subsection*{Acknowledgement}
We gratefully acknowledge a clarifying communication on the sign of
the $q\sq\tilde{\gamma}$ vertex correction by K.~Hikasa. W.B. thanks
the DESY Theory Group for the warm hospitality extended to him during
a visit.

\section*{Appendix}
In this appendix we present the exact supersymmetric one-loop corrections 
to the $q\sq\gl$ and $qqg$ couplings, Eq.~(\ref{gseff}),
discussed in Sect.~\ref{set-up}
for $Q^2 \ll \ms^2 = m_q^2$. The
corrections are given in both the \MS ~and the \DR ~renormalization schemes.

\setlength{\parindent}{0pt}
\subsubsection*{\MS ~Renormalization Scheme:}
$\underline{\text{\emph{vertex corrections}}}$
\begin{eqnarray*}
  (q\sq\gl) & \hat{\delta}_V = & \frac{\as}{4\pi}\,\left\{ C_A\,\left[ 
        -\,\frac{2}{\varepsilon} - \log\left(\frac{Q^2}{\ms^2}\right) 
        + 1 \right] + C_F\,\left[ -\,\frac{6}{\varepsilon} + 6 \right]
        \right\} \\[2mm]
  (qqg) &\delta_V = & \frac{\as}{4\pi}\,C_F\,\left[
        -\,\frac{8}{\varepsilon} + 7 \right]
\end{eqnarray*}
$\underline{\text{\emph{wave-function renormalization}}}$
\begin{eqnarray*}
  (Z_q-1)/2     &=& \frac{\as}{4\pi}\,C_F\,\left[ \frac{4}{\varepsilon} 
                    - \frac{7}{2} \right]\\[2mm]
  (Z_{\sq}-1)/2 &=& \frac{\as}{4\pi}\,C_F\,\left[ \frac{2}{\varepsilon} 
                    - 2 \right] \\[2mm]
  (Z_g-1)/2     &=& \frac{\as}{4\pi}\,\left\{ C_A\,\left[ 
                    -\,\frac{1}{\varepsilon} 
                    - \frac{1}{2}\,\log\left(\frac{Q^2}{\ms^2}\right)
                    + \frac{7}{6} \right]
                    + (n_f+1)\,\frac{1}{\varepsilon} \right\} \\[2mm]
  (Z_{\gl}-1)/2 &=& \frac{\as}{4\pi}\,\left\{ C_A\,\left[ 
                    \frac{1}{\varepsilon} 
                    + \frac{1}{2}\,\log\left(\frac{Q^2}{\ms^2}\right)
                    - \frac{1}{2} \right]
                    + (n_f+1)\,\frac{1}{\varepsilon} \right\}
\end{eqnarray*}
$\underline{\text{\emph{charge renormalization}}}$
\begin{displaymath}
  Z_{g_s} -1 = Z_{\ghat} -1 = \frac{\as}{4\pi}\,\left[
  C_A\,\frac{3}{\varepsilon} - (n_f+1)\,\frac{1}{\varepsilon} \right]
\end{displaymath}

\subsubsection*{\DR ~Renormalization Scheme:}
$\underline{\text{\emph{vertex corrections}}}$
\begin{eqnarray*}
  (q\sq\gl) & \hat{\delta}_V = & \frac{\as}{4\pi}\,\left\{ C_A\,\left[ 
        -\,\frac{2}{\varepsilon} - \log\left(\frac{Q^2}{\ms^2}\right) 
        + 2 \right] + C_F\,\left[ -\,\frac{6}{\varepsilon} + 6 \right]
        \right\} \\[2mm]
  (qqg) & \delta_V = & \frac{\as}{4\pi}\,C_F\,\left[
        -\,\frac{8}{\varepsilon} + 8 \right]
\end{eqnarray*}
$\underline{\text{\emph{wave-function renormalization}}}$
\begin{eqnarray*}
  (Z_q-1)/2     &=& \frac{\as}{4\pi}\,C_F\,\left[ \frac{4}{\varepsilon} 
                    - 4 \right]\\[2mm]
  (Z_{\sq}-1)/2 &=& \frac{\as}{4\pi}\,C_F\,\left[ \frac{2}{\varepsilon} 
                    - 2 \right]\\[2mm]
  (Z_g-1)/2     &=& \frac{\as}{4\pi}\,\left\{ C_A\,\left[ 
                    -\,\frac{1}{\varepsilon} 
                    - \frac{1}{2}\,\log\left(\frac{Q^2}{\ms^2}\right)
                    + 1 \right] + (n_f+1)\,\frac{1}{\varepsilon}
                  \right\} \\[2mm]
  (Z_{\gl}-1)/2 &=& \frac{\as}{4\pi}\,\left\{ C_A\,\left[ 
                    \frac{1}{\varepsilon} 
                    + \frac{1}{2}\,\log\left(\frac{Q^2}{\ms^2}\right)
                    - 1 \right] + (n_f+1)\,\frac{1}{\varepsilon} \right\}
\end{eqnarray*}
$\underline{\text{\emph{charge renormalization}}}$
\begin{displaymath}
 Z_{g_s}-1= Z_{\ghat}-1 =  \frac{\as}{4\pi}\,\left[ C_A\,\frac{3}{\varepsilon} 
                           - (n_f+1)\,\frac{1}{\varepsilon} \right]
\end{displaymath}

\frenchspacing
 \newcommand{\zp}[3]{{\sl Z. Phys.} {\bf #1} (19#2) #3}
 \newcommand{\np}[3]{{\sl Nucl. Phys.} {\bf #1} (19#2)~#3}
 \newcommand{\pl}[3]{{\sl Phys. Lett.} {\bf #1} (19#2) #3}
 \newcommand{\pr}[3]{{\sl Phys. Rev.} {\bf #1} (19#2) #3}
 \newcommand{\prl}[3]{{\sl Phys. Rev. Lett.} {\bf #1} (19#2) #3}
 \newcommand{\fp}[3]{{\sl Fortschr. Phys.} {\bf #1} (19#2) #3}
 \newcommand{\nc}[3]{{\sl Nuovo Cimento} {\bf #1} (19#2) #3}
 \newcommand{\ijmp}[3]{{\sl Int. J. Mod. Phys.} {\bf #1} (19#2) #3}
 \newcommand{\ptp}[3]{{\sl Prog. Theo. Phys.} {\bf #1} (19#2) #3}
 \newcommand{\sjnp}[3]{{\sl Sov. J. Nucl. Phys.} {\bf #1} (19#2) #3}
 \newcommand{\cpc}[3]{{\sl Comp. Phys. Commun.} {\bf #1} (19#2) #3}
 \newcommand{\mpl}[3]{{\sl Mod. Phys. Lett.} {\bf #1} (19#2) #3}
 \newcommand{\cmp}[3]{{\sl Commun. Math. Phys.} {\bf #1} (19#2) #3}
 \newcommand{\jmp}[3]{{\sl J. Math. Phys.} {\bf #1} (19#2) #3}
 \newcommand{\nim}[3]{{\sl Nucl. Instr. Meth.} {\bf #1} (19#2) #3}
 \newcommand{\el}[3]{{\sl Europhysics Letters} {\bf #1} (19#2) #3}
 \newcommand{\ap}[3]{{\sl Ann. of Phys.} {\bf #1} (19#2) #3}
 \newcommand{\jetp}[3]{{\sl JETP} {\bf #1} (19#2) #3}
 \newcommand{\acpp}[3]{{\sl Acta Physica Polonica} {\bf #1} (19#2) #3}
 \newcommand{\vj}[4]{{\sl #1~}{\bf #2} (19#3) #4}
 \newcommand{\ej}[3]{{\bf #1} (19#2) #3}
 \newcommand{\vjs}[2]{{\sl #1~}{\bf #2}}

\pagebreak

\pagebreak

\end{document}